# CMOS-compatible Strain Engineering for High-Performance Monolayer Semiconductor Transistors

Marc Jaikissoon[1], Çağıl Köroğlu[1], Jerry A. Yang[1], Kathryn M. Neilson[1], Krishna C. Saraswat[1,2], Eric Pop[1,2,3,4*]

[1]*Department of Electrical Engineering, Stanford University, Stanford, CA 94305, U.S.A.*
[2]*Department of Materials Science & Engineering, Stanford University, Stanford, CA 94305, U.S.A.*
[3]*Department of Applied Physics, Stanford University, Stanford, CA 94305, U.S.A.*
[4]*Precourt Institute for Energy, Stanford University, Stanford, CA 94305, U.S.A.*

[*]Corresponding author email: epop@stanford.edu

**Strain engineering has played a key role in modern silicon electronics, having been introduced as a mobility booster in the 1990s and commercialized in the early 2000s. Achieving similar advances with two-dimensional (2D) semiconductors in a CMOS (complementary metal oxide semiconductor) compatible manner would radically improve the industrial viability of 2D transistors. Here, we show silicon nitride capping layers can impart strain to monolayer $MoS_2$ transistors on conventional silicon substrates, enhancing their electrical performance with a low thermal budget (350 °C), CMOS-compatible approach. Strained back-gated and dual-gated $MoS_2$ transistors demonstrate median increases up to 60% and 45% in on-state current, respectively. The greatest improvements are found when both transistor channels and contacts are reduced to ~200 nm, reaching saturation currents of 488 µA/µm, higher than any previous reports at such short contact pitch. Simulations reveal that most benefits arise from tensile strain lowering the contact Schottky barriers, and that further reducing device dimensions (including contacts) will continue to offer increased strain and performance improvements.**

Commercial silicon CMOS technology has benefitted from strain boosting of transistor performance for two decades, since the 90 nm technology node[1–5]. Silicon nitride ($SiN_x$) capping layers have been used in silicon nMOS to achieve uniaxial tensile strain and increase electron mobility[6,7], whereas selective SiGe growth in the silicon pMOS source and drain has been used to create uniaxial compressive strain and enhance hole mobility[8]. These improvements are due to changes in the band structure, which lead to a reduction of electron and hole effective masses and scattering rates.

Two-dimensional (2D) semiconductors such as transition metal dichalcogenides (TMDs) have gained recent attention due to their atomically thin nature, which shows potential for transistor scaling[9]. TMDs like monolayer molybdenum disulfide ($MoS_2$) have seen several advances in growth[10,11], doping[12,13], and contact engineering[14–16] for transistor applications. As with silicon, strain engineering has been predicted to modulate the TMD band structure and mobility[17–19], however these effects have only been probed through optical experiments[20,21], or electrical measurements of micron-scale monolayers on bent flexible substrates[22] and bilayers on rigid substrates[23]. To fundamentally enable the integration of TMD transistors with conventional semiconductor technology, strain must be implemented in scalable, CMOS-compatible ways on planar, rigid silicon substrates, ideally boosting the performance of nanoscale monolayer transistors, closer to the fundamental limits of these materials.

**Silicon Nitride Deposition and Back-Gated Devices**

In this work, we apply controllable strain to monolayer $MoS_2$ by depositing $SiN_x$ films with high (yet tunable) intrinsic tensile stress by plasma-enhanced chemical vapor deposition (PE-CVD) at 350 °C. Owing to the low deposition temperature used, this is an attractive back-end-of-line (BEOL) compatible approach, with $SiN_x$ being widely employed in modern semiconductor technology[24]. The stress in these films can be varied from compressive to tensile by varying deposition parameters such as precursor ratio, He gas dilution and pressure[25]. This offers process tunability using a single capping layer, rather than needing separate materials to select between compressive and tensile stress. (Further details of $SiN_x$ deposition and stress tuning are provided in **Supplementary Information Section 1**.)

To understand the strain effects of $SiN_x$ capping on a simplified device geometry, we first examine a conventional back-gated structure, as shown in **Figure 1a**. Monolayer $MoS_2$ is grown by CVD on 90 nm $SiO_2$ on $p^{++}$ Si substrates[26], which also serve as back-gates. The contact metal is 50 nm of Au deposited by electron-beam (e-beam) evaporation at a pressure of $\sim 10^{-8}$ Torr, and further details regarding device fabrication are given in the Methods. **Figure 1b** shows that plasma-induced damage occurs when $SiN_x$ films are directly deposited on $MoS_2$, as indicated by the appearance of the defect-induced LA(M) peak[27] in the Raman spectra of $MoS_2$. Thus, we use a barrier layer of 1.5 nm e-beam evaporated Al followed by 10 nm $AlO_x$ deposited by atomic layer deposition (ALD), which is a common encapsulation[12] for monolayer $MoS_2$. **Figure 1c** shows that when $SiN_x$ is deposited after $AlO_x$, the Raman characteristics of $MoS_2$ do not display visible defect signatures.

We note that Raman analysis cannot be used to accurately estimate strain in such encapsulated $MoS_2$, because doping and plasmon coupling from the $AlO_x$ also affect the E' peak position[28]. Instead, we rely on grazing incidence X-ray diffraction (XRD) measurements for blanket films (**Supplementary Information Section 2**) which indicate that compressive strain is created in $MoS_2$ if a tensile $SiN_x$ layer is deposited on an unpatterned $MoS_2$ film, consistent with previous observations for large-area (micron-scale) capping[20]. However, as we will show below, an important aspect of strain engineering is that the effect of stressor layers depends strongly on the dimensions of the device: the strain can change in both magnitude and sign along the transistor in the presence of metal contacts and gates, especially in nanoscale devices. This effect cannot be mapped by either Raman or XRD analyses due to their large spot sizes (~0.5 μm to several mm) and inability to probe the strain below metal layers. This challenge has been acknowledged in strained-Si technology, where finite-element simulations have been used to understand strain distributions in nanoscale devices, calibrated against transmission electron microscopy with dark-field electron holography[29,30]. To provide such insight, here we use similar finite-element simulations to estimate the strain distributions in our nanoscale 2D devices.

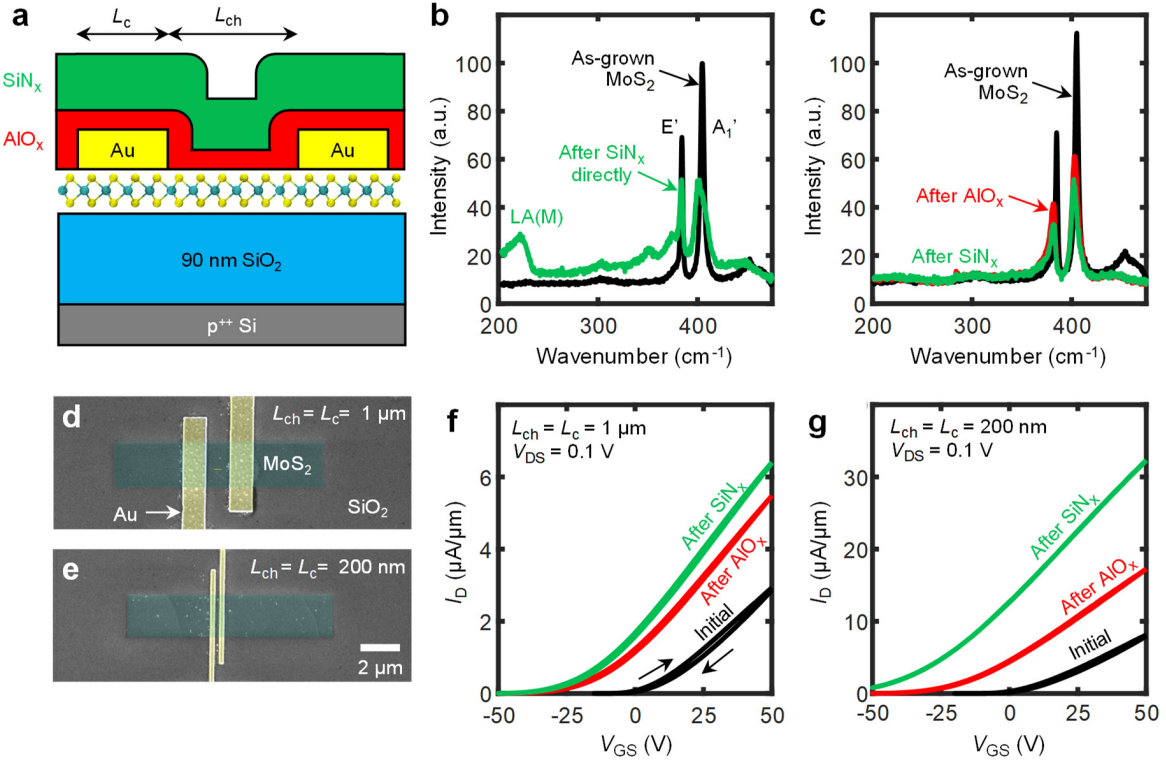

**Fig. 1 | Back-gated transistor characterization. a**, Schematic of back-gated monolayer MoS$_2$ transistor capped with AlO$_x$ and tensile-stressed SiN$_x$. The contact pitch is the sum of channel and contact length, CP = $L_{ch}$ + $L_c$. Raman spectra of monolayer MoS$_2$ before and after direct deposition of SiN$_x$: **b**, without AlO$_x$ buffer layer and **c**, with AlO$_x$ buffer. Top-down, false color scanning electron microscope image of **d**, 'long' device with $L_{ch}$ = $L_c$ = 1 μm and **e**, a 'short' device with $L_{ch}$ = $L_c$ = 200 nm. **f**, Back-gated transfer characteristics of high-stress SiN$_x$-capped MoS$_2$ transistor with 'long' dimensions and **g**, 'short' dimensions. Small arrows mark forward and backward voltage sweeps.

As we mentioned above, capping-layer induced strain in Si nMOS shows a strong dependency on the critical dimensions of the device, with shorter channels experiencing higher strains after capping[31]. For this reason, we investigate the effect of device dimensions by varying the length of the channel ($L_{ch}$) and contacts ($L_c$), using 'long' and 'short' geometries with $L_{ch} = L_c$ of 1 µm and 200 nm, respectively, as shown in **Figure 1d,e**. To understand the effect of capping layers, we performed electrical measurements on the same devices after each step: initial (no capping), after $AlO_x$ is deposited, and after $SiN_x$ is deposited. This allows us to avoid variability which may occur from using different $MoS_2$ growths.

**Figure 1f,g** shows the measured drain current vs. gate voltage ($I_D$-$V_{GS}$) of 'long' and 'short' devices, respectively. In both cases, the $AlO_x$ layer induces *n*-type doping as previously reported[12,32], negatively shifting the threshold voltage ($V_T$) and lowering the contact resistance. On the other hand, capping with $SiN_x$ (75 nm thick, ~600 MPa tensile stress) is noticeably geometry-dependent: the 'long' device has only a small negative $V_T$ shift, while the 'short' device displays both a larger $V_T$ shift, as well as improved transconductance ($g_m = \partial I_D/\partial V_{GS}$) and on-state current ($I_{on}$). While the $V_T$ shift could be attributed to doping from the capping layers, the geometry dependence and improved transconductance point to an origin of this enhancement arising from the stressed $SiN_x$, as we investigate below.

We first confirm that the improvement is reproducible by measuring several (6 to 10) devices with both 'long' and 'short' geometries, as shown with box plots in **Figure 2a,b**. To account for $V_T$ shifting, $I_{on}$ is extracted at a carrier density of $n \approx 8 \times 10^{12}$ cm$^{-2}$. The median $I_{on}$ increases by 14% in 'long' devices after $SiN_x$ capping (**Figure 2a**) and by 60% in 'short' devices (**Figure 2b**). This demonstrates that the effect of high-stress capping is consistent between devices of the same type, with larger increases of $I_{on}$ observed only for the smaller geometry. Other combinations of channel and contact length are shown in **Supplementary Information Section 3**, confirming that both dimensions must be reduced to maximize the $I_{on}$ improvement from this technique. These findings are consistent with those from silicon technology, where greater strain-induced performance ($I_{on}$) is also found in smaller devices[31].

Next, we fabricate control samples by capping with a low-stress (50 to 100 MPa) $SiN_x$ layer of the same thickness and deposition temperature as the high-stress (600 MPa) $SiN_x$ described above. **Figure 2c,d** compares the effect of different stress levels on the relative $I_{on}/I_{on,0}$ at $n \approx 8 \times 10^{12}$ cm$^{-2}$ for both device geometries. (Here, $I_{on,0}$ is the current level after the $AlO_x$ buffer but before $SiN_x$ capping.) For 'long' devices (**Figure 2c**), both low- and high-stress $SiN_x$ layers lead to similar results, with only a small improvement (~10%) of median $I_{on}$. However, for 'short' devices, **Figure 2d** shows that capping with high-stress $SiN_x$ increases the median $I_{on}$ by nearly ~60%, compared to 18% with the low-stress $SiN_x$. This confirms that the large improvement originates from the high tensile stress $SiN_x$, rather than annealing or doping effects which would be similar for both low- and high-stress capping.

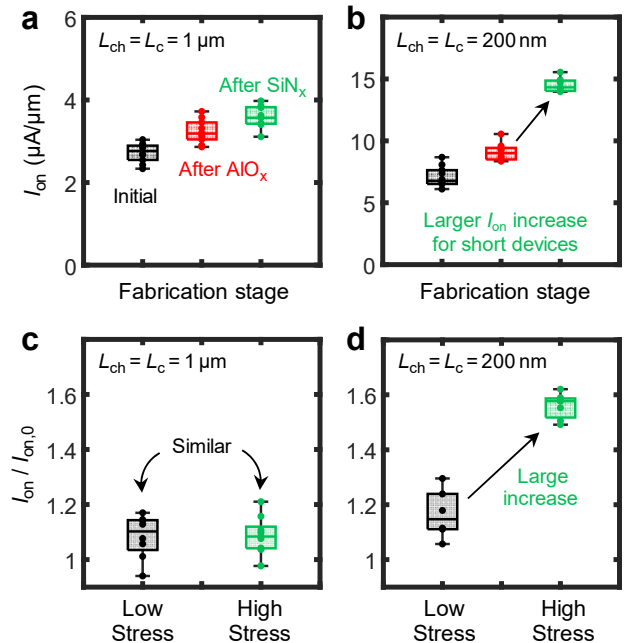

**Fig. 2 | Geometry-dependent device statistics and low-stress control sample.** Box plots of normalized on-state current ($I_{on}$) at $n = 8\times10^{12}$ cm$^{-2}$ for several devices, after each fabrication step. **a**, $L_{ch} = L_c = 1$ μm ('long') and **b**, $L_{ch} = L_c = 200$ nm ('short') devices. Relative improvement in $I_{on}$ after capping with low- (50 to 100 MPa) and high-stress (600 MPa) SiN$_x$ films for **c**, 'long' and **d**, 'short' devices. All measurements are carried out at room temperature and $V_{DS} = 0.1$ V.

## Finite Element Simulations of Strain Profile

To understand the origin of the performance enhancement achieved with high-stress SiN$_x$, we performed finite-element method simulations of such BG devices with various channel and contact lengths to compute the strain distribution. **Figure 3a** shows the 'short' 200-nm device cross-section, with arrows indicating the traction applied by the SiN$_x$ on the underlying AlO$_x$, as well as the resulting displacement field of MoS$_2$. A zoomed-in, exaggerated deformation of the right contact and a color map of the strain field along the channel direction are shown in **Figure 3b**. The tensile SiN$_x$ 'pushes down' on the contact, while simultaneously 'pulling' on its bottom corners[29], like a taut tape simultaneously squeezing and pulling on a small object placed beneath. (See **Supplementary Information Figure 4** for a simple representation of this effect using tape, a drinking straw, and a kitchen sponge.)

Thus, the tensile SiN$_x$ layer imparts a complex, non-uniform strain profile along the MoS$_2$ contact and channel, with uniaxial tensile strain under the contact and compressive strain in the channel, as shown in **Figure 3c**. (More simulation details are included in **Supplementary Information Section 5**.) This

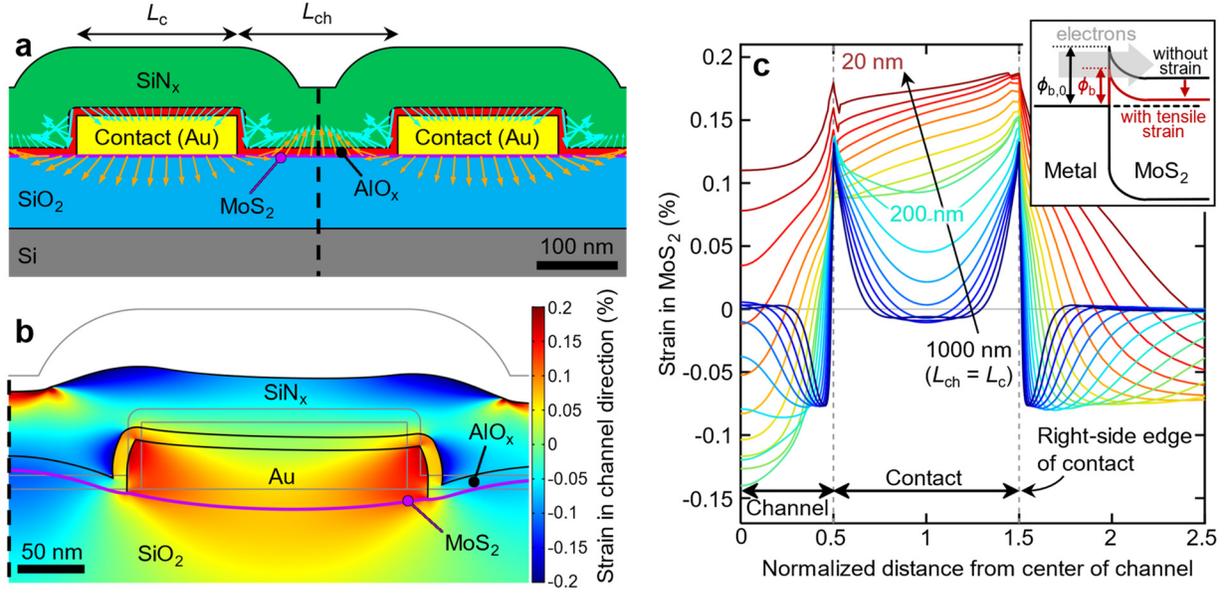

**Fig. 3 | Strain simulations. a,** Cross-section of the 'short' device with $L_{ch} = L_c = 200$ nm. The horizontal magenta line shows the location of the monolayer MoS$_2$. Cyan arrows indicate the traction applied by the SiN$_x$ stressor layer on the underlying AlO$_x$. Orange arrows indicate the displacement of MoS$_2$ after the structure is allowed to relax. The dashed line marks the center of the device, i.e. the symmetry plane. **b,** Zoom into the simulated right-side contact region, with heat map denoting the horizontal strain field (tensile strain is positive). Deformations are exaggerated by a factor of 200. Gray lines are material boundaries prior to deformation. **c,** Strain profile along horizontal direction for devices with $L_{ch} = L_c$ from 1 μm down to 20 nm. Note that distances are normalized by the channel length. Inset: effect of tensile strain on the Schottky barrier at the MoS$_2$ contact.

figure plots the strain along MoS$_2$ for several cases of $L_{ch} = L_c$, from 1 μm to 20 nm. At longer dimensions ($L_{ch} = L_c >$ 150 nm), the tensile strain under the contact is highest near the edges, decaying toward the center of the contact with a characteristic length of ~120 nm, and the compressive strain in the channel is highest near the contact. We estimate tensile strains between 0.1–0.2% near the contact, although we note that factors such as thermal expansion during SiN$_x$ deposition, changes in the elastic modulus of MoS$_2$ due to defects, increases in the SiN$_x$ stress during pre-measurement annealing and slipping between the MoS$_2$ and the substrate can act to increase this strain value in devices. As dimensions are reduced, the tensile strain under the contact increases and becomes more uniform, and the channel strain eventually becomes tensile as well. (Additional trends are explored in more detail in **Supplementary Information Sections 6 and 7**.) Based on these projections, we expect that this technique offers the most benefit at sub-50 nm contact pitches, where both the channel and contact resistances can be greatly improved.

Tensile (compressive) strain distribution in or under mesa-like structures capped with tensile (compressive) stressors have been noted in the silicon literature[29,33] and exploited to enhance the device performance[31]. For a 2D semiconductor like MoS$_2$, tensile strain is expected to lower the K-valley of the conduction band[18,22], bringing it closer to the Fermi level under our contacts and reducing the Schottky barrier height[34] (see **Figure 3c inset**). In addition, the 'downward' pressure exerted by the tensile-strained contacts on the MoS$_2$ (**Supplementary Information Section 8**) could reduce the metal-MoS$_2$ van der Waals gap at the contact[35,36], improving electron tunneling. The corresponding reduction of contact resistance is the likely mechanism for the performance enhancement seen in our devices capped with high tensile stress SiN$_x$, with the greatest enhancement seen in our 'short' (more contact-dominated) devices (**Figure 2d**). Importantly, we also find that additional performance enhancements are possible as the channel and contact lengths are scaled down toward 20 nm, due to increased strain under the contacts as well as the channel going from compressive to tensile strained which could significantly increase the channel mobility[18,22].

To examine the effect of stress on contacts, we estimate the Schottky barrier height of devices capped with high tensile-stressed SiN$_x$. As shown in **Supplementary Information Section 9**, we extract an effective barrier of ~60 meV, which is significantly lower than our control sample measurements and other values from the literature (120-150 meV)[37]. We also perform a pseudo-transfer length method analysis (see **Supplementary Information Section 10**), which confirms that devices with 'short' dimensions exhibit lower contact resistance. These results corroborate the findings of our finite element simulations, showing how strain can be used for CMOS-compatible contact engineering in TMDs.

**Strained Dual-Gate Field-Effect Transistors**

Finally, we extend our strain technique to dual-gated (DG) transistors (schematic in **Figure 4a**), which have a Pd top gate (TG) above the AlO$_x$ encapsulation layer described earlier. As with the earlier BG transistors, we fabricate devices with different values of $L_{ch}$ and $L_c$, and measure their electrical characteristics before and after capping with the SiN$_x$ stress layer above the TG. The scanning electron microscope image of an encapsulated 'short' device ($L_{ch} = L_c = 200$ nm) is shown in **Figure 4b**. For optimal control of the transistor characteristics, we sweep both TG and BG voltages simultaneously, but with different ranges and voltage steps due to the unequal top and back-gate dielectrics.

**Figure 4c** displays electrical measurements of such a 'short' device. After high-tensile stress SiN$_x$ capping, $I_{on}$ increases by 33% (at maximum $V_{BG}$ and $V_{TG}$ applied), with only a small negative $V_T$ shift. Any possible charge transfer doping from the SiN$_x$ encapsulation is effectively blocked by the top-gate, which fully overlaps the channel and contacts. The transconductance (slope, $g_m$) itself increases by 32% after SiN$_x$, underlining that the higher $I_{on}$ is almost entirely due to strain-induced improvements in mobility and contact resistance. Small $V_T$ shifts due to strain are not unexpected (due to changes in the band gap) and have also been observed in silicon technology[38], but can be compensated by gate stack engineering[39]. In addition, the larger band gap of monolayer TMDs (~ 2× larger than Si) implies that any trade-offs in off-state current will be easier to manage than in silicon.

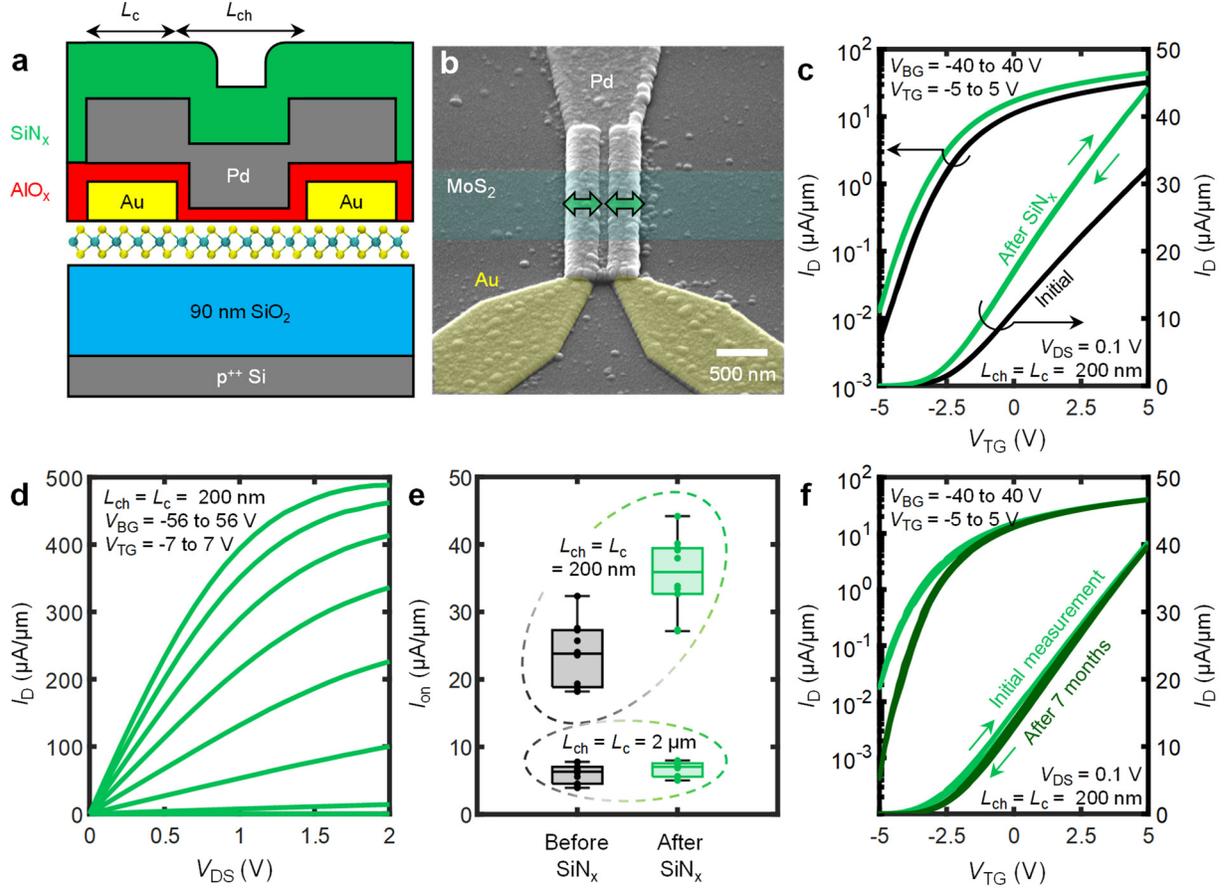

**Fig. 4 | Dual-gated transistors with strain. a**, Schematic of device structure for a dual-gated (TG and BG) monolayer (1L) MoS$_2$ transistor capped with AlO$_x$ and tensile stressed SiN$_x$. **b**, Tilted, colorized scanning electron microscope image of a 'short' device with channel and contact length of 200 nm. The top Pd gate overlaps both source and drain contacts. The green block arrows represent the tensile-stressed SiN$_x$ capping. **c**, Transfer characteristics for a dual-gated high-stress (~800 MPa) SiN$_x$-capped monolayer MoS$_2$ transistor with dimensions $L_{ch} = L_c = 200$ nm. **d**, Output characteristics after SiN$_x$ capping, displaying on-state current of $I_{D,sat} = 488$ µA/µm (with proper saturation at $V_{DS} = 2$ V) when both top and bottom gates are set to high bias. Each voltage step is 16 V on the back gate, and 2 V on the top gate. **e**, Box plots summarizing the change in $I_{on}$ at $V_{DS} = 0.1$ V for DG devices with 'long' and 'short' dimensions after SiN$_x$ capping **f**, Transfer characteristics of a SiN$_x$-capped device after 7 months, showing no visible degradation of the on-state current. Measurements were performed in air, at room temperature. Small arrows mark forward and backward sweeps[40], showing minimal hysteresis over this voltage range, with SiN$_x$ encapsulation. For both **c** and **f**, the $V_{BG}$ and $V_{TG}$ step size have a ratio of 8:1, proportional to their respective voltage sweep ranges.

The output characteristics in **Figure 4d** reveal a saturation current of $I_{D,sat}$ = 488 µA/µm (394 µA/µm) at $V_{DS}$ = 2 V (1 V), the highest reported to date in a 200 nm monolayer MoS$_2$ channel (with 400 nm contact pitch), in a device without otherwise optimized metal contacts or gate dielectrics. This is an important finding, which signifies that strained (but otherwise ordinary) contacts can yield device performance similar to the best Bi or Sb contacts available today[14,15]. The CMOS-compatible strain approach employed in this work is agnostic to the type of contacts, opening the door to future device optimization with more industry-friendly metals.

We summarize measurements of several DG transistors with the 'short' (200 nm) and 'long' (2 µm) geometry in **Figure 4e**. For 'long' devices, $I_{on}$ increases only a few percent after high-stress capping. In contrast, 'short' devices display a large median $I_{on}$ increase of 45%. This effect is reproducible across all properly-strained transistors and confirms our BG device findings (**Figures 2** and **3**), that strain boosts the performance of devices with smaller channel length and contact pitch (here, 200 nm and 400 nm, respectively). On a separate test chip, we find that 'short' DG devices which used MoS$_2$ grown at lower temperature[11] (thus, more weakly adhered to the substrate) suffered partial delamination due to SiN$_x$ strain and displayed no improvement in their $I_{on}$ (see **Supplementary Information Section 11**), again confirming a strain-related source of improvement in our 'short' well-adhered devices. For properly-strained devices, our DG simulations in **Supplementary Information Section 6** show similar strain distributions as in BG transistors, indicating the performance increase is primarily due to tensile contact strain, and projecting further enhancement in smaller devices. These simulations also provide design guidelines on how channel and contact dimensions affect strain, independently.

Finally, we test the stability of our method by measuring transistor characteristics over time, as shown in **Figure 4f**. After 7 months, the device shows no degradation of on-state current even when these measurements are performed in air, illustrating that SiN$_x$ also offers robust encapsulation, which is known for Si transistors, as a diffusion barrier to moisture and gases[24,41]. Our CMOS-compatible strain technique can also be applied to other TMDs which are expected to benefit from tensile strain[17], such as monolayer WSe$_2$ (see **Supplementary Information Section 12**). We anticipate that this approach will offer further improvements to 2D transistors with even shorter critical dimensions, paving the way for the implementation of CMOS-compatible strain in high-performance TMDs.

**Conclusions**

We reported the first CMOS-compatible approach to impart strain to 2D semiconductor transistors, using low-temperature, tensile-stressed silicon nitride capping layers. This improves the performance of monolayer MoS$_2$ transistors up to 60%, reaching saturation current of 488 µA/µm, a record at a contact pitch of just 400 nm (channel + contact). Simulations reveal that strain is expected to have even greater benefits in future transistors with smaller contact pitches, e.g. sub-50 nm. The results of this study are essential for integrating 2D semiconductors into future electronics, and this work is also likely to motivate future exploration of strain engineering in other next-generation semiconductors.

**Methods**

**Device Fabrication.** Monolayer MoS$_2$ was synthesized by chemical vapor deposition (CVD) at 750 °C directly onto thermally-grown SiO$_2$ (90 nm) on 1.5 cm × 2 cm p$^{++}$ silicon substrates, as previously reported[26]. For back-gated (BG) devices, MoS$_2$ was used directly on the growth substrates, where the p$^{++}$ Si also serves as BG. Electron-beam (e-beam) lithography (Raith VOYAGER) was employed for each patterning step, using poly(methyl methacrylate) (PMMA) as e-beam resist to minimize potential delamination from aqueous developers used in photolithography. First, alignment marks (2 nm Ti/40 nm Au) were patterned and deposited by lift-off after e-beam evaporation. Discrete single-crystal MoS$_2$ triangles were identified under a microscope and lithography masks were designed such that each device was fabricated within a single triangle of MoS$_2$, in regions with minimal overgrowth. The channel dimensions were then defined using XeF$_2$ etching for 90 s at 3 Torr (Xactix e-1). Large probing pads (20 nm SiO$_2$/2 nm Ti/20 nm Pt) were then patterned and deposited via lift-off, with the evaporated SiO$_2$ layer serving to reduce potential leakage to the back-gate. Finally, source and drain device contacts (50 nm Au without an adhesion layer[16]) were used to connect the MoS$_2$ channels with the probing pads, patterned and deposited by lift-off using e-beam evaporation at < 5×10$^{-8}$ Torr and a rate of 0.5 Å/s. All lift-off processes were performed in acetone for a minimum of 2 hours.

Next, e-beam evaporation was used to deposit a 1.5 nm Al film[12] onto of finished BG devices, with Al oxidizing in air to sub-stoichiometric AlO$_x$. Subsequently, 10 nm Al$_2$O$_3$ is deposited by atomic layer deposition at 200 °C (Cambridge Nanotech Savannah S200). This serves as the barrier layer prior to SiN$_x$ capping for BG devices, and as the top-gate insulator for DG devices. At this stage, the Al$_2$O$_3$ is wet-etched (JT-Baker, Aluminum etch) over the contact pads after optical lithography. BG devices are then encapsulated by SiN$_x$, which is removed only over the contact pads after optical lithography and CF$_4$ plasma etching (Samco PC300). For DG devices, e-beam lithography is used to define the gate metal (50 nm Pd) via lift-off, followed by SiN$_x$ deposition and contact access as described above.

All transistors were tested with a Keithley 4200 semiconductor parameter analyzer. BG transistors are measured in a Janis ST-500 probe station under ~10$^{-5}$ Torr vacuum after *in-situ* annealing at 250 °C

11for 2 hours. DG transistors are measured in air due to minimal hysteresis after encapsulation. All measurements are done at room temperature unless otherwise stated.

**Silicon Nitride Deposition.** Silicon nitride ($SiN_x$) was deposited using plasma-enhanced CVD (PECVD) in a PlasmaTherm Shuttlelock PECVD system using $SiH_4$ (5% in He) and $NH_3$ gases. Several recipes were developed in order to control the resultant film stress, and it was found that the major process variable which can be used to control stress is the ratio of $NH_3$ to $SiH_4$. At higher ratios of $NH_3$:$SiH_4 > 1$ film stress becomes tensile, whereas lower ratios promote compressive stress (see **Supplementary Information Section 1** for more details). Additionally, the amount of He in the process chamber can be used to further tune the stress if needed. For the transistors in this work, the deposition temperature was set to 350 °C although this can be reduced as low as 130 °C, both temperatures being back-end of line (BEOL) compatible. Deposition powers were tuned from 20 to 100 W, with process pressures between 1 to 2 Torr resulting in deposition rates between 10 to 15 nm/min. Films were characterized using ellipsometry to precisely fit both refractive index and thickness in order to obtain accurate information about the deposition, which is necessary for accurate stress measurement. Film stress was measured on reference 4" silicon wafers using a Flexus 2320 Stress Tester via radius of curvature measurements before and after deposition.

**Acknowledgements**

This work was performed in part at the Stanford Nanofabrication Facility (SNF) and the Stanford Nano Shared Facilities (SNSF), which are supported by the National Science Foundation (NSF) under award ECCS-2026822. This work was partly supported by the Stanford SystemX Alliance, Samsung Global

Research Outreach (GRO) program, and Intel Corporation. M.J. and K.M.N acknowledge support from the Stanford Graduate Fellowship. C.K. and E.P. acknowledge the support of the ASCENT and SUPREME JUMP Centers, both Semiconductor Research Corporation (SRC) programs sponsored by DARPA. J.A.Y. acknowledges support from NSF Graduate Research Fellowship. The authors wish to thank James P. McVittie and Muyu Xue for fruitful discussions.

**Author contributions**

M.J., E.P. and K.C.S. conceived the work. M.J. performed the $MoS_2$ CVD growth, $SiN_x$ recipe development, device fabrication, optical characterization, electrical measurement and scanning electron microscopy. C.K. contributed all numerical simulations of strain profiles. K.M.N. performed $WSe_2$ CVD growth, and atomic layer deposition with J.A.Y. M.J. analyzed all data and wrote the manuscript with help from C.K. and E.P. All authors have given approval to the final version of the manuscript.

**Data availability**

The data that support the plots within this paper and other findings of this study are available from the corresponding author upon reasonable request.

**Competing financial interests**

The authors declare no competing financial interests.

## Supplementary Information

# CMOS-compatible Strain Engineering for High-Performance Monolayer Semiconductor Transistors

Marc Jaikissoon[1], Çağıl Köroğlu[1], Jerry A. Yang[1], Kathryn M. Neilson[1], Krishna C. Saraswat[1,2], Eric Pop[1,2,3,4*]

[1]*Department of Electrical Engineering, Stanford University, Stanford, CA 94305, U.S.A.*
[2]*Department of Materials Science & Engineering, Stanford University, Stanford, CA 94305, U.S.A.*
[3]*Department of Applied Physics, Stanford University, Stanford, CA 94305, U.S.A.*
[4]*Precourt Institute for Energy, Stanford University, Stanford, CA 94305, U.S.A.*

[*]Corresponding author email: epop@stanford.edu

**1. Silicon Nitride Film Stress Measurement**

We benchmark our $SiN_x$ deposition by measuring film stress on reference 4" silicon wafers as described in the Methods section. The Stoney equation $r = (E_s t_s^2)/[6(1-\nu_s)\sigma_f t_f]$ was used, where $r$ is the radius of curvature of the sample measured by laser deflection, $\sigma_f$ is the desired film stress, $t_f$ is the film thickness measured by ellipsometry, $\nu_s$ is the Poisson's ratio of the substrate, $t_s$ is the substrate thickness and $E_s$ is the Young's modulus of the substrate. **Supplementary Figure 1** shows the measured film stress of several $SiN_x$ films deposited with various ratios of $NH_3$:$SiH_4$, demonstrating the tunable nature of stress from compressive (≈ -600 MPa) to tensile (≈ 800 MPa).

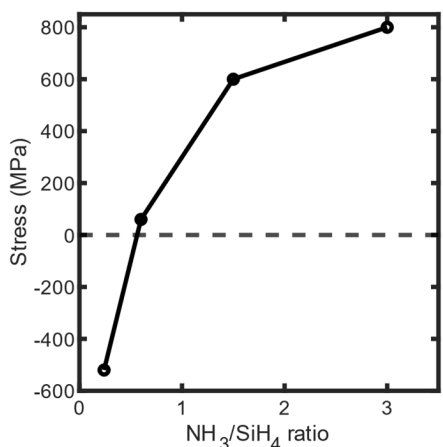

**Supplementary Figure 1 | $SiN_x$ stress measurement**. Measured film stress in $SiN_x$ deposited by PE-CVD at 350 °C as a function of $NH_3$:$SiH_4$ precursor ratio.

## 2. Raman and XRD Analysis of Strain in As-Grown and Encapsulated MoS₂ Films

We acquired Raman spectra at each stage of encapsulation of a blanket MoS$_2$ film (on SiO$_2$/Si), and the fitted E' vs. A$_1$' peak positions[1] are summarized in **Supplementary Figure 2a** below. The AlO$_x$ capping (1.5 nm Al seed + 10 nm Al$_2$O$_3$ by ALD) causes a large, nearly ≈ 4 cm$^{-1}$, redshift of the E' peak of MoS$_2$. This would appear to imply a tensile strain > 0.75%, however we have found in previous work[2] that Raman analysis greatly overestimates MoS$_2$ strain under AlO$_x$ encapsulation (as compared to accurate X-ray diffraction measurements). The apparent shift of the E' peak is instead caused by doping and plasmon coupling of the MoS$_2$ with the AlO$_x$ encapsulation layer.

We could still gain some insight into the effect of blanket tensile SiN$_x$ deposition (on AlO$_x$/MoS$_2$) by comparing peak positions before and after SiN$_x$ capping (green dots), which indicates slight compression ($\Delta\varepsilon \approx -0.15\%$) relative to AlO$_x$/MoS$_2$ data points (red dots), as well as increased electron doping ($\Delta n \approx 5\times10^{12}$ cm$^{-2}$). This has been well-documented in the literature[3], and indicates that compressive strain is created in MoS$_2$ if a tensile SiN$_x$ layer is deposited on an *unpatterned* film. (The tensile SiN$_x$ contracts to relieve its built-in stress, in the process compressing the MoS$_2$ underneath.)

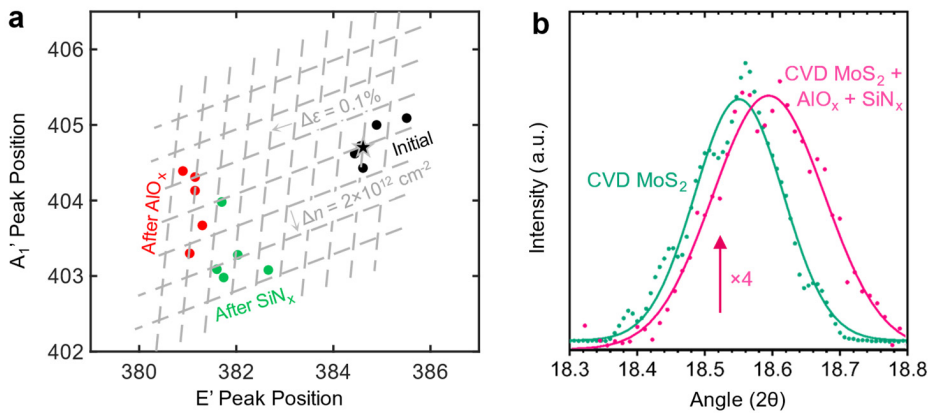

**Supplementary Figure 2 | Estimating strain by Raman and X-Ray Diffraction on unpatterned MoS₂. a,** Raman E' vs. A$_1$' peak positions acquired at several spots on unpatterned, as-grown MoS$_2$ (on SiO$_2$/Si) before and after encapsulation with AlO$_x$ and then SiN$_x$. Dashed lines mark expected strain ($\varepsilon$) and electron density ($n$) changes[1]. We emphasize that the apparent strain after AlO$_x$ coverage (~0.75%) is greatly overestimated by Raman analysis[2]. **b,** Grazing-incidence X-ray diffraction spectra acquired on as-grown MoS$_2$ before and after capping with AlO$_x$ and high-tensile stress SiN$_x$.

To probe strain more accurately in unpatterned MoS$_2$ films, we used grazing incidence X-ray diffraction measurements with a synchrotron X-ray source on as-grown MoS$_2$, before and after capping with AlO$_x$ and high-tensile stress SiN$_x$, as shown in **Supplementary Figure 2b**. We observe an increase of 0.0422° of the in-plane (10) peak position of MoS$_2$, corresponding to a compressive biaxial strain of -0.23% after capping, again consistent with expectations for *unpatterned* films.

However, we note that both the Raman laser and X-ray spot sizes are large (~0.5 μm to several mm), which makes strain mapping difficult using these techniques in any sub-micron devices. Additionally, neither approach allows characterization of the full strain profile across a transistor including the regions of MoS$_2$ under the (metal) contacts and top gate, indicating the need for more sophisticated metrologies, such as transmission electron microscopy (TEM)[4,5], to measure strain in nanoscale devices. Therefore, we used finite element simulations (main text **Figure 3** and **Supplementary Information Sections 5-8**) to provide insight into the strain profiles and distributions in capped MoS$_2$, particularly within our nanoscale transistors (where strain is uniaxial, due to the presence of contacts). Such simulations have also been adopted for nanoscale silicon devices[4,5], and have been found in agreement with TEM-based metrology.

## 3. Other Back-gated Channel and Contact Combinations

In addition to using 'long' and 'short' geometries (channel and contact length $L_{ch} = L_c = 1$ µm and 200 nm, respectively) mentioned in the main text, we also tested other combinations of $L_{ch}$ and $L_c$. The drain current vs. gate voltage ($I_D$-$V_{GS}$) of devices with $L_{ch} = 1$ µm, $L_c = 200$ nm and $L_{ch} = 200$ nm, $L_c = 1$ µm are shown in **Supplementary Figure 3a,b**. Both cases demonstrate negative threshold voltage ($V_T$) shifts as reported for the devices in the main text after AlO$_x$ and SiN$_x$ capping, as well as improvements to transconductance ($g_m = \partial I_D/\partial V_{GS}$). Normalizing by carrier density to account for $V_T$ shifts, box plots of $I_{on}$ extracted at a carrier density of $n \approx 8\times10^{12}$ cm$^{-2}$ are shown for $L_{ch} = 1$ µm, $L_c = 200$ nm and $L_{ch} = 200$ nm, $L_c = 1$ µm in **Supplementary Figure 3c,d**. We observe that $I_{on}$ only increases by up to 27% for these combinations, indicating that larger improvements are only possible when both $L_{ch}$ and $L_c$ are reduced, as in the main text.

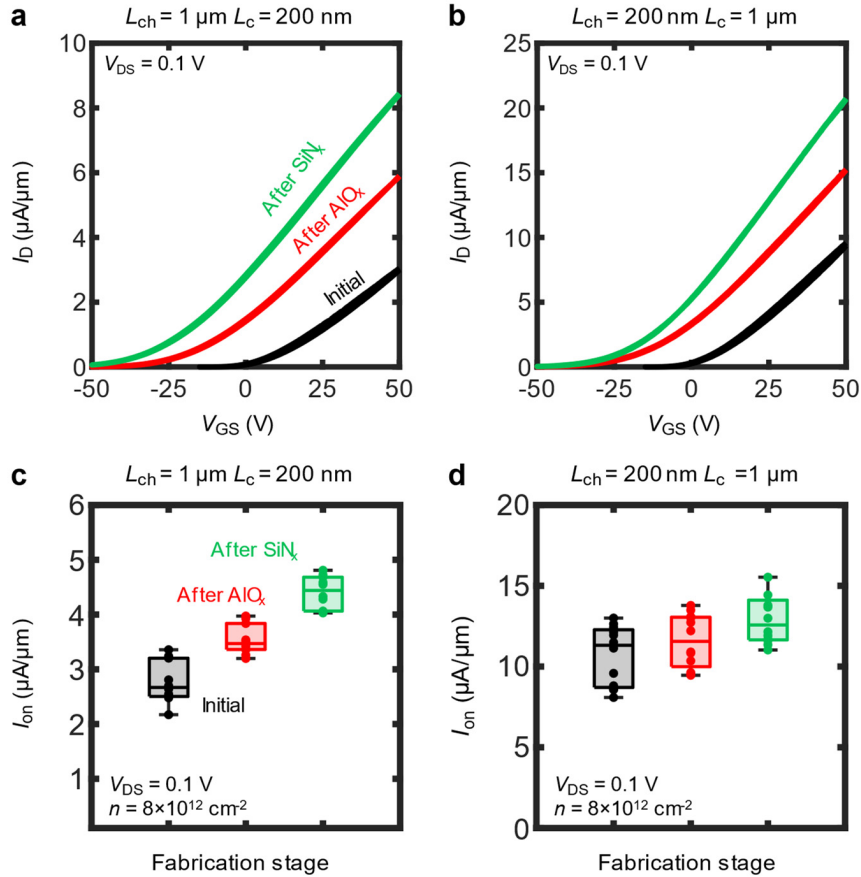

**Supplementary Figure 3 | Characterization of other back-gated transistor geometries.** Back-gated transfer characteristics of high-stress SiN$_x$-capped MoS$_2$ transistor with **a,** $L_{ch} = 1$ µm, $L_c = 200$ nm and **b,** $L_{ch} = 200$ nm, $L_c = 1$ µm. Box plots of normalized on-state current ($I_{on}$) at $n = 8\times10^{12}$ cm$^{-2}$ for several devices, after each fabrication step. **c,** $L_{ch} = 1$ µm, $L_c = 200$ nm and **d,** $L_{ch} = 200$ nm, $L_c = 1$ µm. All measurements are carried out at room temperature and $V_{DS} = 0.1$ V.

## 4. Macroscopic Visual Model of Deformation due to Tensile-Stressed Capping Layer

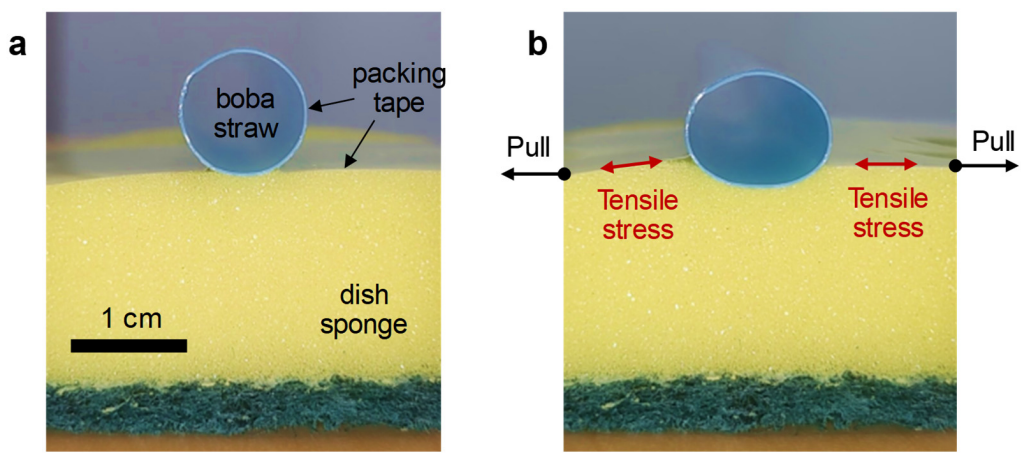

**Supplementary Figure 4 | Visualization of contact electrode deformation using everyday objects. a,** The cross-section of a simple macroscopic visual analogy for the contact geometry of a back-gated $MoS_2$ transistor. Here, a plastic straw (from boba tea) stands in for the contract electrode, a dish sponge for the materials under the electrode ($MoS_2/SiO_2/Si$), and a piece of transparent packing tape (covering the top of the sponge and wrapping as an "Ω" around the straw) for the $SiN_x$ film, initially unstressed. **b,** Contact deformation visualized when the duct tape is laterally tensile-stressed by pulling the tape outward from the sides, similar to the exaggerated simulated device deformation shown in **Figure 3b** of the main text.

## 5. Stress Simulations: Additional Information

Two-dimensional (2D) and three-dimensional (3D) stress simulations were performed for back-gated (BG) and dual-gated (DG) transistors, assuming linear elasticity. We confirmed through 3D simulations that the MoS$_2$ strain in the transistor width direction is small (i.e. MoS$_2$ strain is essentially uniaxial), as illustrated by **Supplementary Figure 5**. Consequently, 2D simulations are sufficient to accurately capture the uniaxial stress and strain distributions in our devices, and the results presented in this work were obtained through 2D simulations. After multi-scale simulations of the entire sample (including the transistor and the entire silicon substrate) confirmed that strains due to substrate bowing were negligible, later simulations used a smaller domain around the BG transistor with a fixed boundary condition at the bottom of a thinner section of substrate, with no appreciable errors in stress and strain distributions.

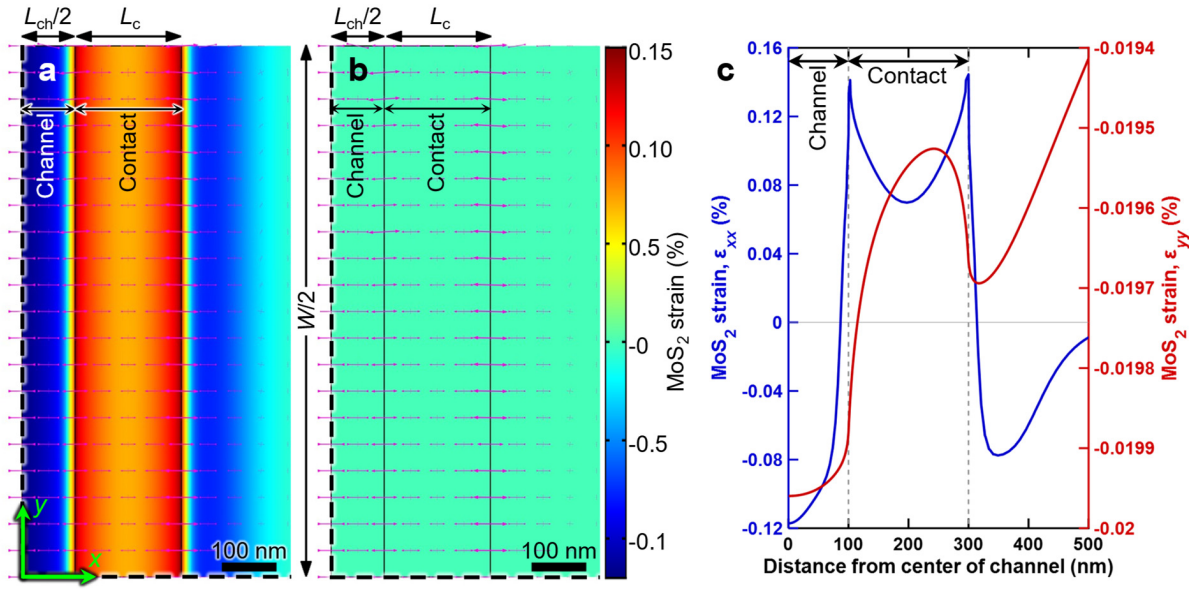

**Supplementary Figure 5 | In-plane strain distribution across the device. a, b,** The distributions of lengthwise in-plane strain ($\varepsilon_{xx}$ given in **a**) and widthwise in-plane strain ($\varepsilon_{yy}$ given in **b**) in MoS$_2$ in a BG transistor (with 600 MPa tensile-stressed SiN$_x$ capping) for $L_{ch} = L_c = 200$ nm, as viewed from above. The bottom left corner corresponds to the center of the device, with the dashed lines indicating the two symmetry planes. The magenta arrows indicate the principal strain directions and values at each point, showing that the strain is predominantly in the direction of current flow (i.e. along the *x*-axis), and hence uniaxial. **c,** MoS$_2$ strains $\varepsilon_{xx}$ (left axis) and $\varepsilon_{yy}$ (right axis) along positive *x*-axis of the same device showing the widthwise strain is small, with a nearly uniform compressive strain < 0.02% in magnitude.

An isotropic "initial stress" (the stress before the geometry is allowed to relax) of 600 MPa was assumed in the SiN$_x$ capping layer. The isotropic elastic properties assumed for the materials other than MoS$_2$ are summarized in **Supplementary Table 1**. In contrast, MoS$_2$ is only transversely isotropic (i.e. isotropic in-plane), and thus is described by an anisotropic stiffness tensor. The elastic properties were taken from Li *et al.*[6], and can be summarized as $E_{xx} = E_{yy} = 265$ GPa, $E_{zz} = 100$ GPa, $G_{xz} = G_{yz} = 50$ GPa, and $v_{xy} = v_{yx} = v_{xz} = v_{yz} = 0.25$. Here, $x$ and $y$ correspond to the in-plane directions and $z$ to the cross-plane direction, $E$ denotes Young's modulus, $G$ denotes shear modulus and $v_{ij}$ denotes Poisson's ratio for loading along $i$ and transverse direction $j$. The remaining elastic properties can be determined from these, e.g. $v_{zx} = (E_{zz}/E_{xx})v_{xz}$ and $G_{xy} = E_{xx}/[2(1 + v_{xy})]$.

**Supplementary Table 1:** Young's moduli and Poisson's ratios assumed for materials except for MoS$_2$.

|  | Si | SiO$_2$ | AlO$_x$ | SiN$_x$ | Au | Pd |
|---|---|---|---|---|---|---|
| **Young's modulus (GPa)** | 170 | 70 | 400 | 250 | 70 | 73 |
| **Poisson's ratio** | 0.28 | 0.17 | 0.22 | 0.23 | 0.44 | 0.44 |

The MoS$_2$ grown by CVD (at 750 °C) directly on SiO$_2$ is tensile-stressed, which has been attributed to the mismatch in coefficients of thermal expansion[7–9] $\alpha_{MoS2}$ and $\alpha_{SiO2}$. To model this observation, we take $\alpha_{MoS2} = 7\times10^{-6}$ K$^{-1}$ and $\alpha_{SiO2} = 2\times10^{-6}$ K$^{-1}$, yielding an initial thermally-induced strain of $(\alpha_{MoS2} - \alpha_{SiO2})(T_{growth} - T_{ambient}) \cong 0.37\%$ in MoS$_2$, which represents the tensile strain in a planar MoS$_2$ film as-grown, relative to relaxed MoS$_2$. The in-plane strains reported in this work are all relative to MoS$_2$ as-grown: the residual MoS$_2$ strain (assumed 0.37% here) should be added to these values to obtain strains relative to relaxed MoS$_2$. We note that the built-in tensile stress in MoS$_2$ has negligible effect on the in-plane strains in MoS$_2$ in the relaxed structure, because MoS$_2$ is so thin and its film force (stress times thickness) is small compared to that effected by the stress in SiN$_x$. In other words, the MoS$_2$ strain is dictated mainly by the adjacent materials, and ultimately caused by the tensile stress in SiN$_x$.

We also note that it is possible for there to be some amount of slipping between MoS$_2$ and adjacent materials due to poor adhesion. While the quantitative details of this process are scarce and slipping is not included in the results we present, if MoS$_2$ is allowed to slip freely on the underlying SiO$_2$, simulations predict that the strains in both the channel and under the contacts increase by up to ~50%, and the strain peaks near the contact edges become more "rounded." However, the trends in the main text and **Supplementary Information Section 6** otherwise stay the same.

We note that all strain simulations in the subsequent sections assume a tensile-stressed SiN$_x$ capping layer (600 MPa, as measured in our experiments) on top of either a DG or BG transistor geometry.

## 6. Strain Projections for Channel and Contact Scaling

We carried out additional simulations to study the impact of the $SiN_x$ stressor on $MoS_2$ strain distributions in BG and DG transistors with different channel and contact lengths. Because our DG transistor geometry is similar to that of a typical top-gated (TG) transistor except for the conductive substrate, the corresponding conclusions apply equally well to TG devices (with no BG). The metrics we focus on are the average channel strain (in-plane, along the channel direction), average strain in $MoS_2$ under the contacts, and $MoS_2$ under the edge of the contact. The latter parameter is quantified as the average strain in the first 30 nm (or $L_c$, whichever is smaller) of $MoS_2$ under the contact, on the channel side. This parameter is relevant because the current under a contact is only distributed within about a few transfer lengths $L_T$ of the contact edge, which is typically[10,11] on the order of tens of nanometers in good contacts with $MoS_2$.

**Supplementary Figure 6** shows the variation of the in-plane strain as a function of $L_{ch} = L_c$, for the BG (**Supplementary Figure 6a,b**) and DG (**Supplementary Figure 6c,d**) devices. **Supplementary Figure 6a** shows that in a BG transistor, the strain under contact is highest and approximately equal at the two edges. However, according to **Supplementary Figure 6c**, a DG transistor has lower strain under the "inner" contact edge than under the "outer" edge. This is because the nitride stressor only directly covers the outer side of the contacts in the DG device and not both sides like it does in the BG geometry. Consequently, as can be seen in **Supplementary Figure 6b,d**, the DG transistor has lower average tensile strain under the contacts for $L_{ch} > 100$ nm, but in shorter devices the DG transistor has higher tensile strain under the contacts as well as in the channel. Reducing the channel and contact lengths toward 20 nm in both BG and DG transistors increases the tensile strain under the contacts substantially, suggesting further improvements of contact resistance (with strain) are possible. Moreover, the tensile-strained channel at shorter channel lengths could lead to increased mobility[12,13], and thus, further performance enhancement. The kink seen in **Supplementary Figure 6b,d** at $L_{ch} = 160$ nm is the result of the fact that for $L_{ch} < 160$ nm, the curved sections of the nitride film above the channel, where it smoothly conforms around the edges of the contact (in the BG device) or the top gate (in the DG device), begin to merge.

We also varied $L_{ch}$ and $L_c$ separately to study their individual effects on the strain distribution in $MoS_2$, the results are given in **Supplementary Figure 7**. According to **Supplementary Figure 7a**, reducing $L_{ch}$ below 160 nm in a BG transistor reduces the tensile strain under the contacts, especially for long contacts. This happens because in the limit of very short channels, the contacts effectively "merge", and the strain under the inner contact edge is lower for the same reason the strain in under the middle of the contact is lower than at the edges. Reducing $L_{ch}$ also increases the channel strain significantly, turning it tensile for $L_{ch}$ close to 20 nm. According to **Supplementary Figure 7b**, reducing $L_c$ has a similar effect on the BG transistor channel strain, and also increases both the edge and average contact strains as it is reduced beyond 160 nm. **Supplementary Figure 7c** shows that in a DG transistor, reducing $L_{ch}$ similarly increases the tensile channel strain, although its effect on contact strain is less pronounced than in the BG case. Finally, a comparison of **Supplementary Figure 7d** to **Supplementary Figure 7b** reveals that the effect of reducing DG transistor contact lengths is similar to the BG transistor, increasing strain both in the channel and under contacts.

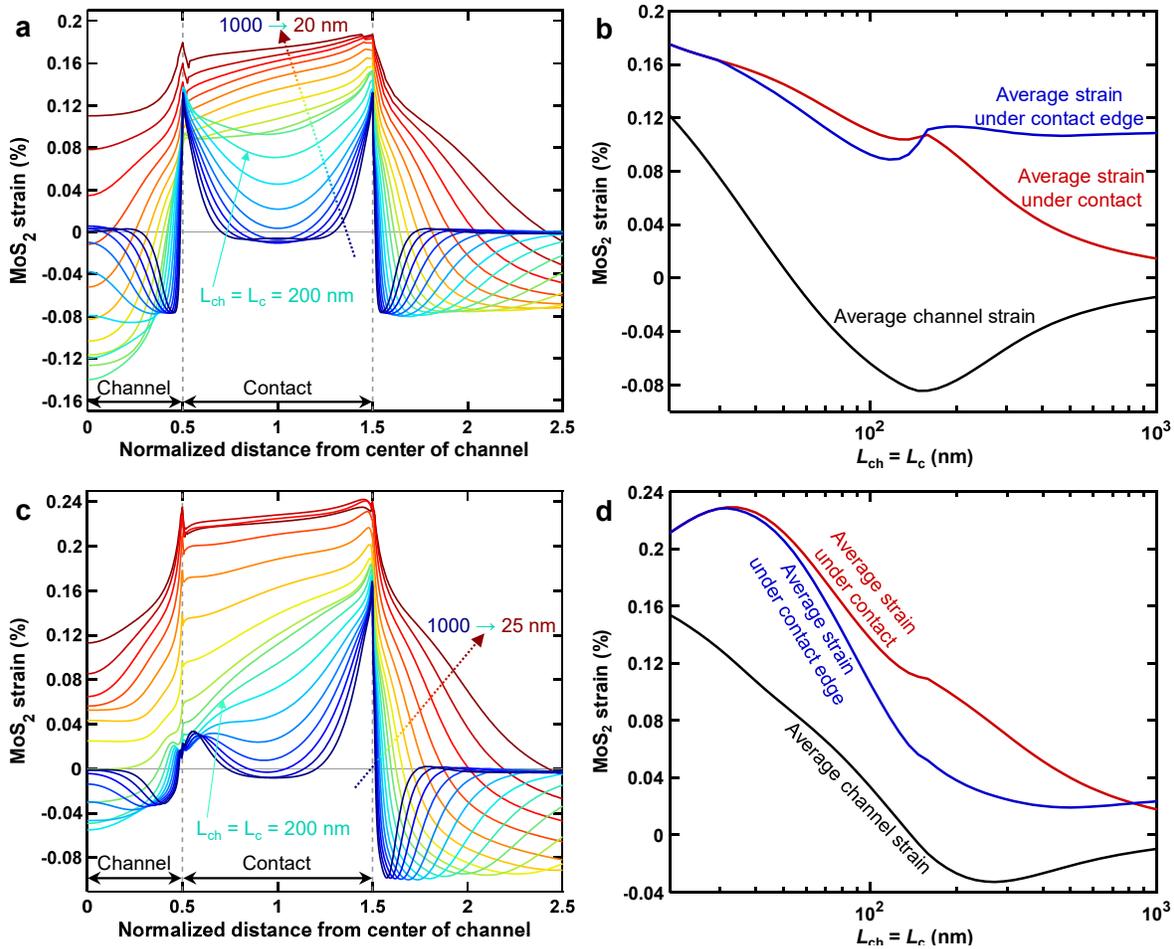

**Supplementary Figure 6 | Stress simulations and projections. a,** The strain profiles along MoS$_2$ in a BG transistor (with 600 MPa tensile-stressed SiN$_x$ capping), normalized by $L_{ch}$, for $L_{ch} = L_c$ varied from 1000 nm down to 20 nm (logarithmically spaced). **b,** The corresponding average in-plane strains along MoS$_2$ as a function of $L_{ch} = L_c$. **c,** The normalized strain profiles along MoS$_2$ in a DG transistor, for $L_{ch} = L_c$ varied from 1000 nm down to 25 nm (logarithmically spaced). **c,** Average in-plane strains in MoS$_2$ in a DG transistor, as a function of $L_{ch} = L_c$. **d,** The corresponding average in-plane strains along MoS$_2$ as a function of $L_{ch} = L_c$.

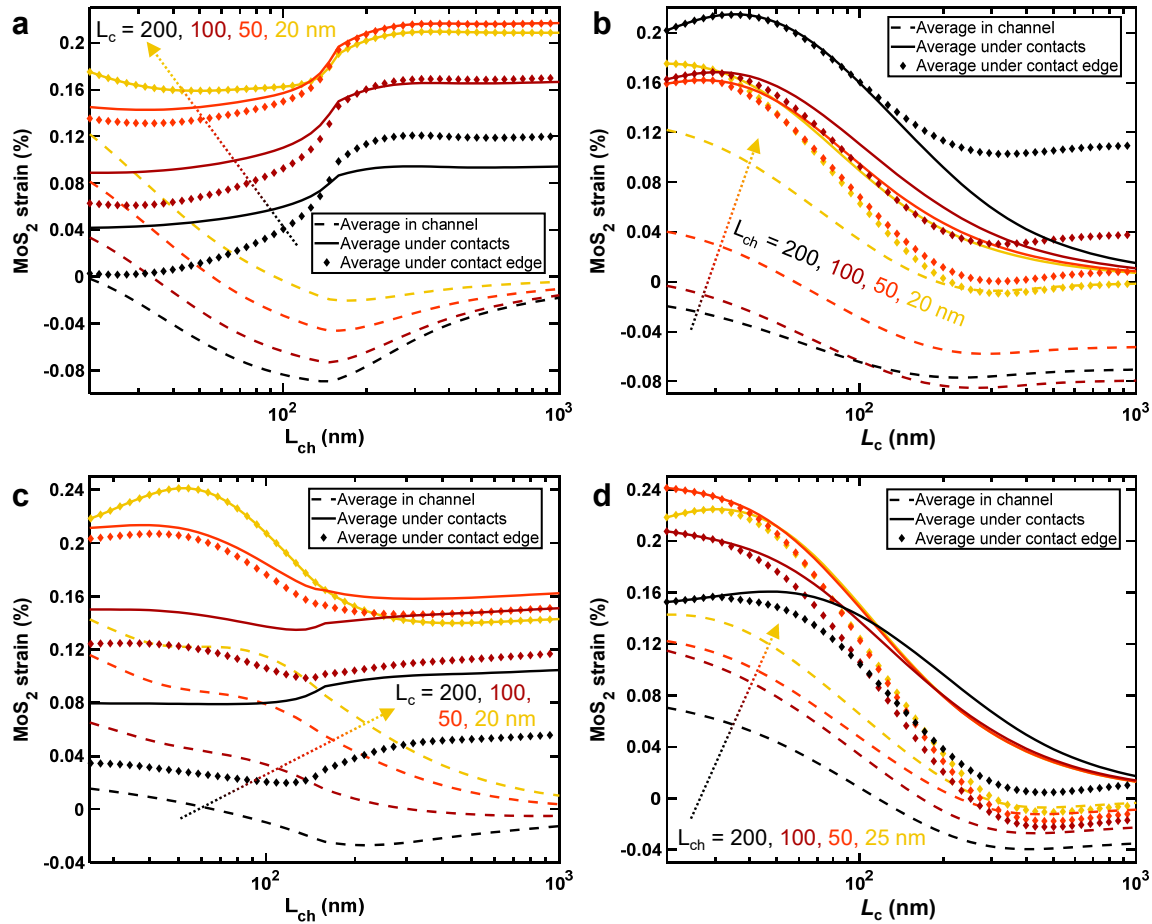

**Supplementary Figure 7 | Strain projections for channel and contact scaling.** Average in-plane strains in a back-gated (BG) MoS$_2$ transistor with a tensile-stressed SiN$_x$ capping layer (600 MPa) when **a,** $L_{ch}$ is varied with $L_c$ fixed, and **b,** $L_c$ is varied with $L_{ch}$ fixed. Average in-plane strains in MoS$_2$ in a DG transistor when **c,** $L_{ch}$ is varied with $L_c$ fixed, and **d,** $L_c$ is varied with $L_{ch}$ fixed.

To summarize, in both BG and DG (or TG) transistors, shorter channels are expected to put tensile strain on the channel, while shorter contacts increase the tensile strain under the contacts as well as in the channel. Both of these effects are expected to increase performance in smaller devices, due to tensile strain under the contacts reducing contact resistance[14], and tensile strain in the channel increasing mobility[12,13].

## 7. Impact of Al$_2$O$_3$ Barrier Thickness on Strain Distribution

We have carried out simulations to understand the effect of the Al$_2$O$_3$ barrier layer thickness on the strain distribution in the BG transistor, presented in **Supplementary Figure 8**. It is seen that the MoS$_2$ strain in both the channel and the contact regions decrease steadily as Al$_2$O$_3$ is made thicker. As such, to further improve contact resistance via strain, there is possibility of enhancing the strain under contact edges by up to 50% by making the Al$_2$O$_3$ thinner than what we have used in this work (10 nm), or possibly even eliminating it.

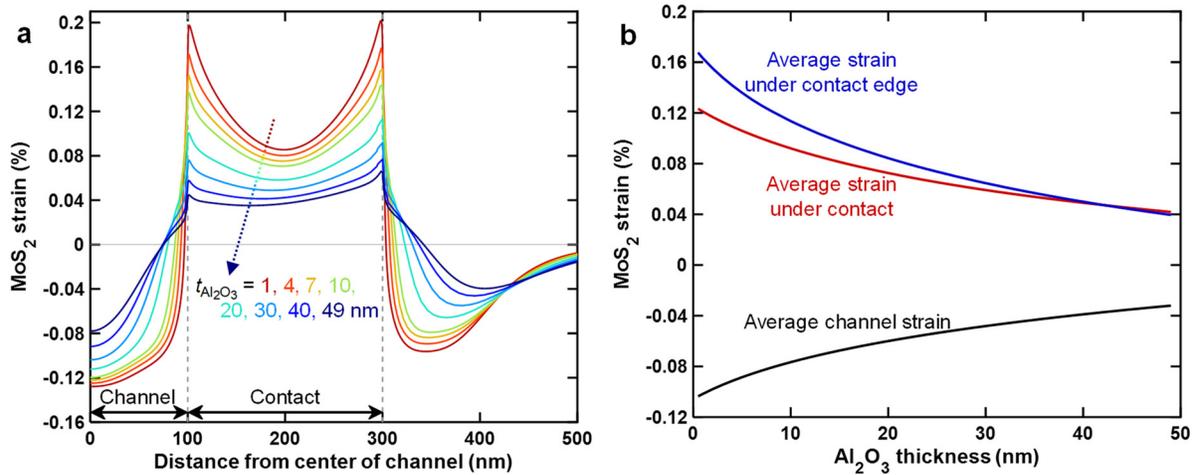

**Supplementary Figure 8 | Dependence of strain profile on Al$_2$O$_3$ barrier layer thickness. a,** In-plane strain distributions in a BG transistor with $L_{ch} = L_c = 200$ nm for several Al$_2$O$_3$ thicknesses ($t_{Al_2O_3}$). **b,** Average in-plane strains in MoS$_2$ in a BG transistor (with 600 MPa tensile-stressed SiN$_x$ capping) as a function of Al$_2$O$_3$ thickness. The blue curve corresponds to the average strain in the first 30 nm of MoS$_2$ (a typical contact transfer length) under the contacts.

## 8. Cross-Plane Stress in MoS$_2$ due to "Downward Pressure" on Contacts

The tensile-stressed SiN$_x$ capping layer pushes down on the contacts due to its tendency to contract, as visualized in **Figure 3a,b** of the main text, as well as **Supplementary Figure 4**, which may reduce the thickness of the van der Waals gap (an electron tunneling barrier) between the Au contacts and MoS$_2$. It is then possible that the vertical contact-MoS$_2$ pressure (i.e. the cross-plane MoS$_2$ stress) due to this effect (in addition to the in-plane tensile strain of MoS$_2$ under the contacts) contributes to the contact resistance improvement we observe in SiN$_x$-capped devices, as the reduction in MoS$_2$ contact resistance with applied pressure has been observed experimentally[15,16].

To better understand the role of the SiN$_x$ in vertically compressing the contacts, we plot the cross-plane stress in BG (**Supplementary Figure 9a,b**) and DG (**Supplementary Figure 9c,d**) MoS$_2$ devices, as a function of $L_{ch} = L_c$ from 1000 nm down to ~20 nm. For both BG and DG devices, the compressive cross-plane stress at the contact is close to about 50 MPa for 50 nm < $L_{ch} = L_c$ < 200 nm. Based on measurements of Chen et al.[15] this would correspond to a contact resistance reduction of ~12%.

Another interesting feature observed in **Supplementary Figure 9b,d** is that while the vertical stress in the channel is moderately tensile for $L_{ch} = L_c$ > ~100 nm, scaling the devices down to $L_{ch} = L_c$ ~20 nm results in a sizable compressive vertical channel stress, up to ~200 MPa. In other words, the MoS$_2$ channel is "vertically squeezed" by the contacts at the shortest channel and contact dimensions (here ~20 nm). This vertical compression of MoS$_2$ has a similar effect on the conduction band structure as lateral tensile strain[17], i.e. lowering the K valley and raising the Q valley (thus expected to reduce intervalley scattering and improve mobility[12,13]), and has also been reported experimentally to improve in-plane conduction[15]. Therefore, the vertical compression of the MoS$_2$ channel could enable further enhancement of device performance at the smallest dimensions, in addition to the effect of in-plane tensile strain discussed in **Supplementary Information Section 6**.

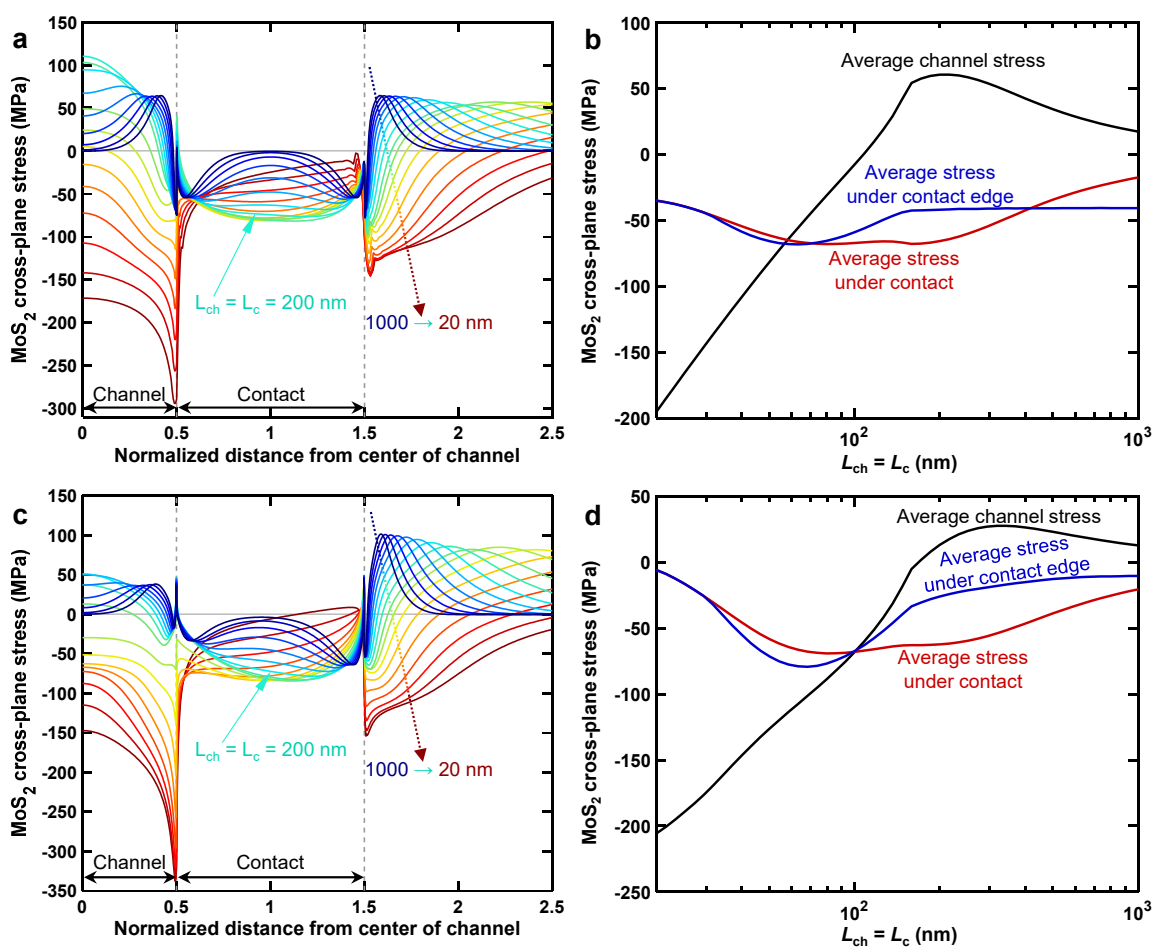

**Supplementary Figure 9 | Cross-plane stress in MoS₂. a,** Normalized cross-plane stress profiles along MoS$_2$ in back-gated (BG) transistors (with 600 MPa tensile-stressed SiN$_x$ capping), normalized by $L_{ch}$, for $L_{ch} = L_c$ from 1000 nm down to 20 nm (logarithmically spaced). **b,** Corresponding average cross-plane stresses along BG MoS$_2$ as a function of $L_{ch} = L_c$. "Average stress under contact edge" refers to the cross-plane stress *averaged* along the first 30 nm of contact length (30 nm is a typical value of contact transfer length). **c,** Normalized cross-plane stress profiles along MoS$_2$ in dual-gated (DG) transistors, for $L_{ch} = L_c$ from 1000 nm down to 20 nm (logarithmically spaced). **d,** Corresponding average cross-plane stresses in MoS$_2$ in DG transistors, as a function of $L_{ch} = L_c$.

## 9. Schottky Barrier Height Measurement

To further investigate the behavior at the contacts, we estimate the Schottky barrier height (SBH) based on the thermionic emission current dictated by the equation:

$$I_\mathrm{D} = A_{2D}^* T^{3/2} \exp\left(-\frac{q\Phi_{SBH}}{k_\mathrm{B}T}\right)\left[1 - \exp\left(-\frac{qV}{k_\mathrm{B}T}\right)\right],$$

where $A_{2D}^*$ is the 2D-equivalent Richardson constant, $T$ is the temperature, $q$ is the elementary charge, $k_\mathrm{B}$ is Boltzmann's constant, $V$ is the applied voltage, and $\Phi_{SBH}$ is the Schottky barrier height[18,19]. **Supplementary Figure 10a-c** shows the results of temperature-dependent measurements to extract the SBH of a strained device having $L_\mathrm{ch}$ = 1 μm and $L_\mathrm{c}$ = 200 nm at $V_\mathrm{DS}$ = 0.1 V, while Supplementary **Figure 10d-f** shows the results of measurements performed on an uncapped control sample with Au contacts having $L_\mathrm{ch}$ = 2 μm and $L_\mathrm{c}$ = 200 nm.

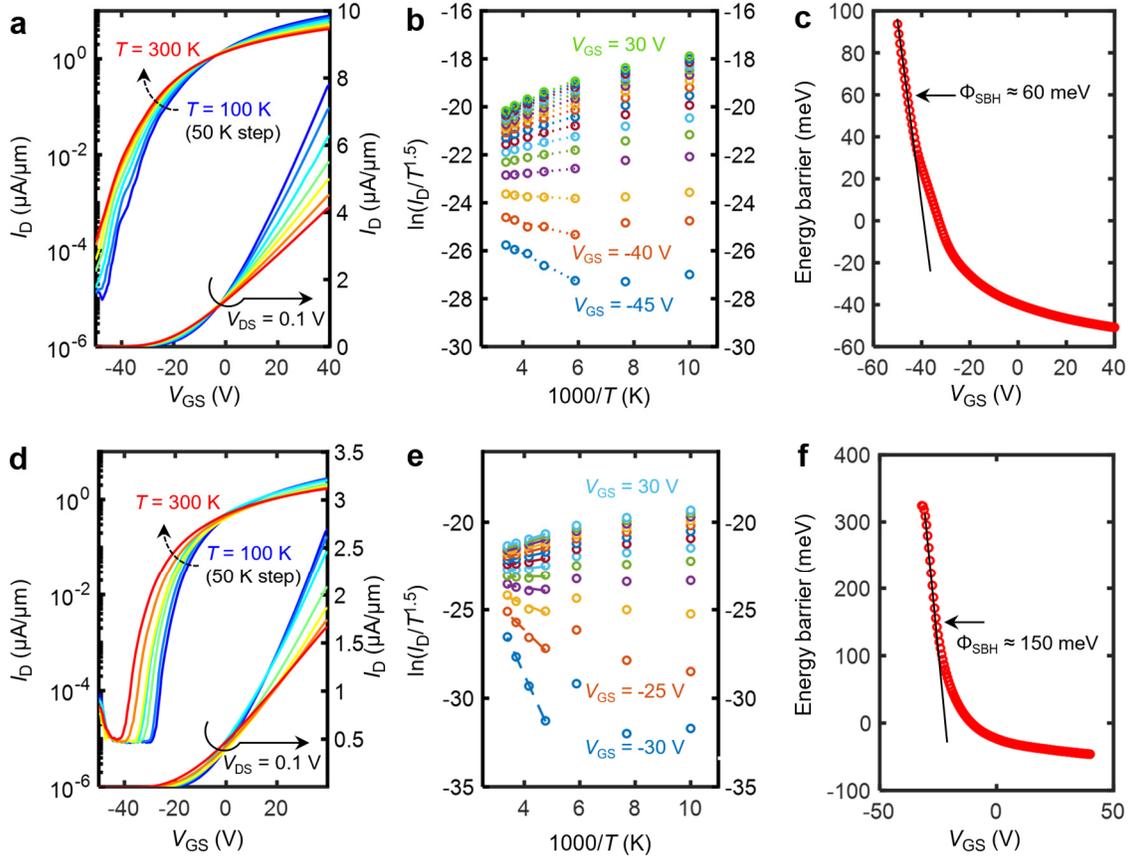

**Supplementary Figure 10 | Schottky barrier height extraction.** Measurements, **a-c,** (top row) of a high tensile-stress SiN$_x$-capped device and **d-f,** (bottom row) an uncapped control device with Au contacts. **a, d**, Temperature-dependent transfer characteristics measured between $T$ = 100 and 300 K at $V_\mathrm{DS}$ = 0.1 V. **b, e**, Arrhenius plots measured at various gate voltages. **c, f**, Estimates of the effective electron Schottky barrier height.

**Supplementary Figure 10a** shows the device $I_D$-$V_{GS}$ characteristics measured from $T$ = 100 to 300 K. From this, the slope of the Arrhenius plot of $\ln(I_D/T^{3/2})$ vs $1000/T$ can be constructed at each value of $V_{GS}$, with the slope giving the effective barrier height at that particular bias as shown in **Supplementary Figure 10b**. Finally, the estimated energy barrier is plotted vs. $V_{GS}$ in **Supplementary Figure 10c**, with the true value of Schottky barrier height (SBH) being determined as the effective barrier height at the flat band voltage indicated by the point above which the effective barrier height starts to deviate from a linear dependence of the gate voltage. We extract a SBH of ~60 meV for the stressed device. The results of SBH measurement on a control device (without $SiN_x$ capping, but otherwise identical) are shown in **Supplementary Figure 10d-f**, indicating a barrier of ~ 150 meV.

## 10. Pseudo-Transfer Length Method Analysis

We estimate contact resistance ($R_C$) and effective electron mobility ($\mu_{eff}$) using pseudo-transfer length method[20–22] (TLM) measurements. We call these "pseudo" TLM measurements, because we fit the median resistance ($R_{tot}$) vs. channel length ($L_{ch}$) for *all* devices we have, rather than choosing a single TLM structure[10] with shared contacts and a larger range of $L_{ch}$, fabricated in a single region of $MoS_2$.

**Supplementary Figure 11a** displays $R_{tot} = L_{ch}R_{sh} + 2R_C$ vs. $L_{ch}$, where $R_{sh}$ is the channel sheet resistance, for devices with 'long' contacts ($L_c = 1$ µm) at various stages of capping. To account for threshold voltage ($V_T$) variation, we normalize all devices to the same maximum gate overdrive ($V_{ov} = V_{GS} - V_T$) using linear extrapolation to estimate $V_T$. The *y*-intercept of the linear fit allows us to extract $2R_C$, which is plotted vs. overdrive voltage in **Supplementary Figure 11b**. For devices with 'long' contacts, the extracted $R_C$ remains similar at different stages of capping. This reflects the lower average stress across the devices with long contacts. Additionally, the slope of the fit allows us to estimate the effective mobility, $\mu_{eff} = (qnR_{sh})^{-1}$ where $q$ is the elementary charge and $n = C_{ox}(V_{GS}-V_T-V_{DS}/2)/q$ is the electron concentration per unit area. **Supplementary Figure 11c** shows the extracted $\mu_{eff}$ vs. overdrive voltage for devices with 'long' contacts, indicating a small improvement after SiN$_x$ capping. We caution against relying too strongly on TLM mobility estimates, because the strain distribution (between the various channel lengths in the TLM) is non-uniform. In addition, we cannot be certain that the apparent mobility increase is entirely due to strain; other contributions could come from dielectric screening and the additional thermal annealing seen by the AlO$_x$/SiN$_x$ capped samples. For these reasons we have put more emphasis on changes in transistor current density ($I_D$) in the main text. The current density (at a given voltage, e.g. 1 V) is also ultimately what impacts the circuit performance of a transistor.

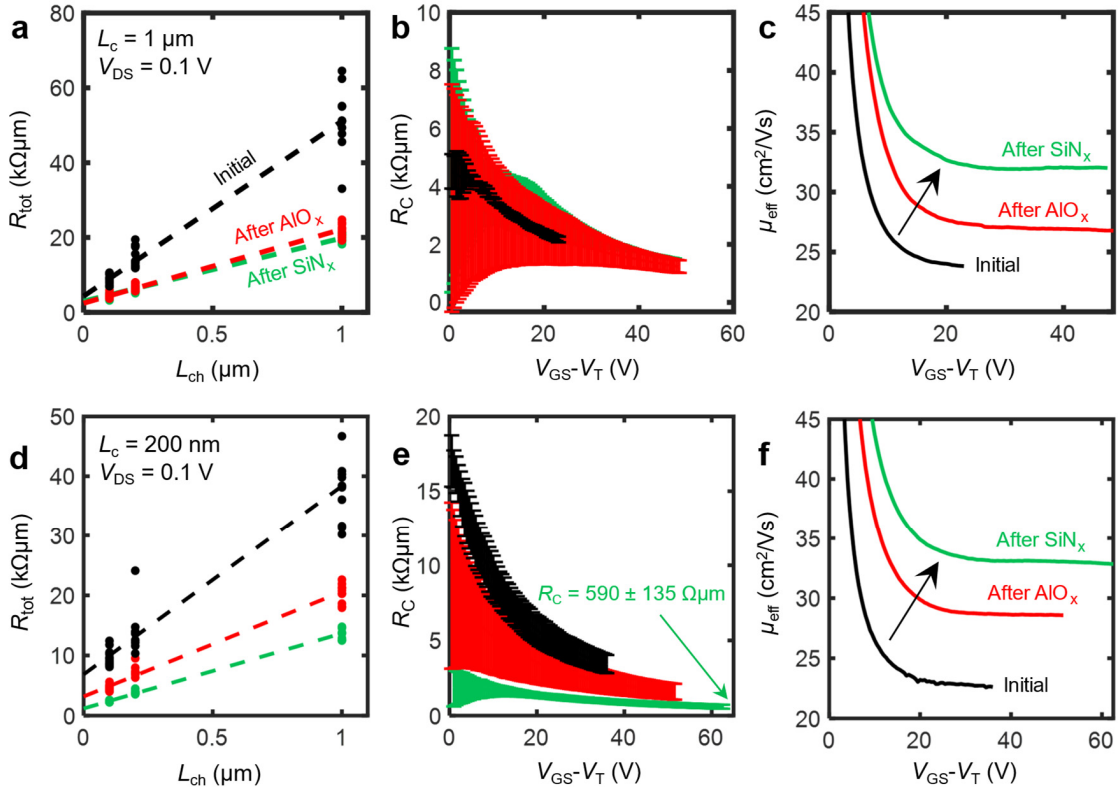

**Supplementary Figure 11 | Pseudo-transfer length method (TLM) analysis.** Comparison of high tensile-stress $SiN_x$-capped devices fabricated with (**a-c**) long, 1 µm and (**d-f**) short, 200 nm contacts. **a, d,** $R_{tot}$ vs. $L_{ch}$ at maximum gate overdrive ($V_{GS} - V_T$) for $V_{DS} = 0.1$ V, at different stages of capping. **b, e,** Extracted $R_C$ vs. gate overdrive voltage. We observe reduced $R_C$ after $SiN_x$ capping only for the short contacts. **c, f,** Estimated effective mobility $\mu_{eff}$ vs. gate overdrive voltage.

We repeat this analysis for devices with 'short' contacts ($L_c$ = 200 nm) in **Supplementary Figure 11d**. After SiN$_x$ capping, the extracted $R_C$ is now significantly reduced to 590 ± 135 Ω·µm as shown in **Supplementary Figure 11e**, demonstrating that high-stress capping has a larger effect on the shorter contacts. The extracted mobility trend is similar in **Supplementary Figure 11f**, which is expected because the channel dimensions used are consistent with those in **Supplementary Figure 11a-c**.

## 11. Measurement of Transistors That Suffered Stress-Induced Delamination

Here, we present additional evidence for strain-related improvements in our transistors by illustrating the effect of strain release/delamination. **Supplementary Figure 11a** shows a focused ion-beam scanning electron microscope (FIB-SEM) cross sectional image for one of our dual-gated 'short' ($L_{ch} = L_c$ = 200 nm) devices, such as the one in main text Figure 3, where we see conformal coverage of $SiN_x$ around the device. By comparison, **Supplementary Figure 11b** shows the cross-section of another 'short' device on a separate chip which experienced delamination after tensile $SiN_x$ capping; this occurred because here we used $MoS_2$ grown at a lower temperature, which is more weakly adhered to the substrate[23]. **Supplementary Figure 11c** (same as main text Figure 4c) and **Figure 11d** compare the measured $I_D$ vs. $V_{GS}$ of a typical well-adhered 'short' transistor with that of a similar device which undergoes delamination, respectively. The former experiences 33% improvement in $I_{on}$ after capping while the latter displays no observable improvement in $I_{on}$. This provides further confirmation that strain transfer is the main source of the improvements observed in our 'short' well-adhered devices.

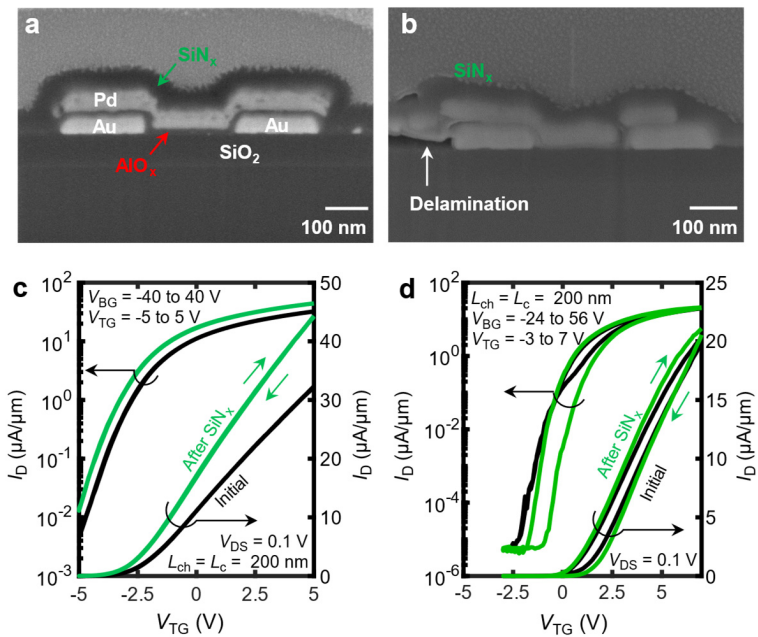

**Supplementary Figure 11** | FIB-SEM cross-sections of **(a)** typical well-adhered dual-gated $MoS_2$ transistor capped with tensile $SiN_x$ and **(b)** a transistor on a separate chip which experienced stress release/delamination after $SiN_x$ capping. These are 'short' devices with $L_{ch} = L_c$ = 200 nm. Measured transfer characteristics for **(c)** a typical well-adhered 'short' transistor and **(d)** a similar transistor which experienced stress release. We observe no visible improvement in the on-state current when stress release occurs, indicating that strain (rather than encapsulation or annealing) is the source of improvement in our well-adhered devices. All measurements are at room temperature and $V_{DS}$ = 0.1 V.

## 12. Strained Dual-Gated WSe₂ Transistors

Monolayer WSe$_2$ is another TMD which is predicted to benefit from the application of uniaxial tensile strain. To further verify the effectiveness of our strain technique, we fabricated dual-gated (DG) transistors using CVD-grown monolayer WSe$_2$ using the approach described in the main text, with the only process difference being the use of 10 nm HfO$_x$ deposited by atomic layer deposition at 200 °C serving as the top gate dielectric, instead of 10 nm AlO$_x$. The $I_D$-$V_{GS}$ of DG WSe$_2$ transistors measured before and after capping with high tensile-stressed SiN$_x$ are shown for 'long' ($L_{ch} = L_c = 1$ μm) and 'short' ($L_{ch} = L_c = 200$ nm) devices in **Supplementary Figure 12a,b**, respectively.

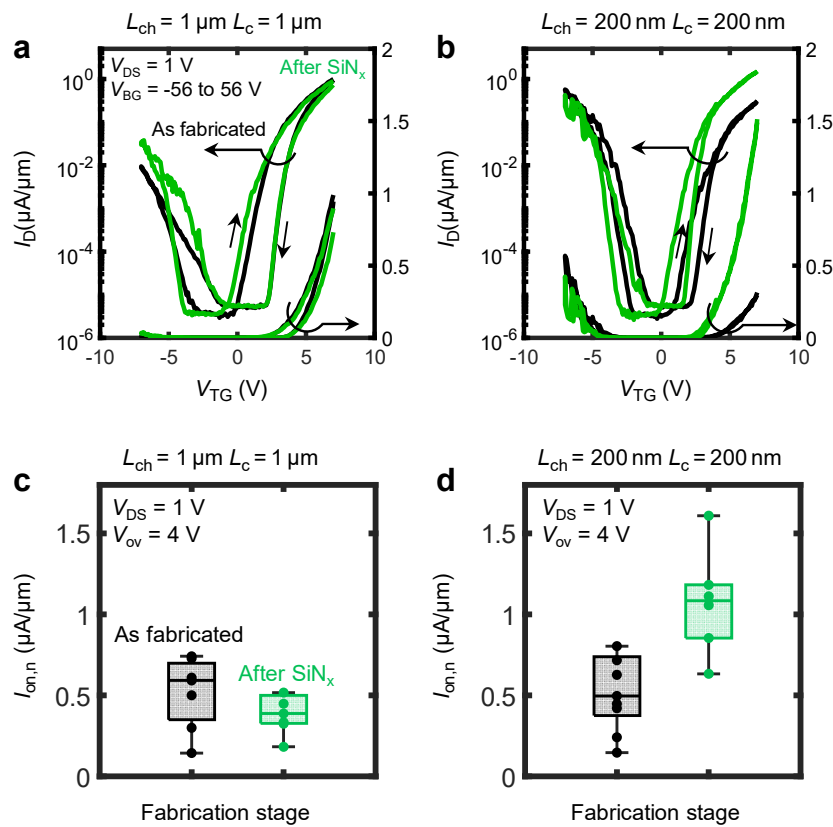

**Supplementary Figure 12 | Strained DG WSe₂ transistors.** Dual-gated transfer characteristics of high-stress SiN$_x$-capped WSe$_2$ transistor with **a**, 'long' ($L_{ch} = L_c = 1$ μm) and **b**, 'short' ($L_{ch} = L_c = 200$ nm) dimensions. Small arrows mark forward and backward sweeps. Relative improvement in n-branch $I_{on}$ at fixed overdrive after capping with high-stress SiN$_x$ films (green) for **c**, 'long' and **d**, 'short' devices. All measurements are carried out at room temperature and $V_{DS} = 1$ V.

We compare the n-branch $I_{on}$ (at fixed $V_{ov} = V_{TG} - V_T > 0$) of each case in **Supplementary Figure 12c,d**. Similar to the trends observed in the case of monolayer MoS$_2$ (main text **Figure 4c,e**), the larger improvement of $I_{on}$ is observed for the devices with the shortest channel and contact lengths.